\documentclass[twocolumn,showpacs,preprintnumbers,amsmath,amssymb,prl]{revtex4}

\usepackage{amsmath}
\usepackage{graphicx}
\usepackage{color}
\usepackage{upgreek}

\begin{document}

\title{Observation of the Eckhaus Instability in Whispering-Gallery Mode Resonators}

\author{Dami\`a Gomila$^{1}$}
\thanks{damia@ifisc.uib-csic.es} 
\author{Pedro Parra-Rivas$^{2,3}$} 
\author{Pere Colet$^1$} 
\author{Aur\'elien Coillet$^{4,5}$} 
\author{Guoping Lin$^{4,6}$} 
\author{Thomas Daugey$^{4}$} 
\author{Souleymane Diallo$^{4}$} 
\author{Jean-Marc Merolla$^{4}$} 
\author{Yanne K. Chembo$^{4,7}$}
\thanks{ykchembo@umd.edu}

\affiliation{$^1$Instituto de F\'isica Interdisciplinar y Sistemas Complejos, IFISC (CSIC-UIB), 
                 Campus Universitat de les Illes Balears, E-07122 Palma de Mallorca, Spain \\
             $^2$Laboratory of Dynamics in Biological Systems, KU Leuven Department of Cellular 
                 and Molecular Medicine, University of Leuven, B-3000 Leuven, Belgium\\
             $^3$Service OPERA-photonique, Universite libre de Bruxelles (ULB),  
                 50 Avenue F. D. Roosevelt, CP194/5 B-1050 Bruxelles, Belgium \\
             $^4$FEMTO-ST Institute, Univ. Bourgogne Franche-Comt\'e, CNRS,  
                 15B Avenue des Montboucons, 25030 Besan\c con cedex, France \\
             $^5$Laboratoire Interdisciplinaire Carnot de Bourgogne, Univ. Bourgogne-Franche-Comt\'e, CNRS, 
                 9 Avenue A. Savary, 21078 Dijon, France \\
             $^6$MOE Key Laboratory of Fundamental Quantities Measurement, School of Physics, 
                 Huazhong University of Science and Technology, Wuhan 430074, China \\
             $^7$University of Maryland, Department of Electrical and Computer Engineering,
                 \& Institute for Research in Electronics and Applied Physics (IREAP) \\
                  8279 Paint Branch Dr, College Park MD 20742, USA}

\date{\today}

\begin{abstract}
The Eckhaus instability is a secondary instability of nonlinear spatiotemporal patterns in which high-wavenumber periodic solutions become unstable against small-wavenumber perturbations. 
We show in this letter that this instability can { take place} in Kerr combs generated with ultra-high $Q$ whispering-gallery mode resonators. {In our experiment, sub-critical Turing patterns (rolls) undergo Eckhaus instabilities upon changes in the laser detuning 
leading to cracking patterns with long-lived transients.}
In the spectral domain, this results in a metastable Kerr comb dynamics with a timescale that can be larger than one minute.
This ultra-slow timescale is at least seven orders of magnitude larger than the intracavity photon lifetime, 
and is in sharp contrast with all the transient behaviors reported so far in cavity nonlinear optics,
that are typically only few photon lifetimes long (i.~e., in the ps to $\upmu$s range).
We show that this phenomenology is well explained by the Lugiato-Lefever model, as 
the result of an Eckhaus instability. Our theoretical analysis is found to be in excellent agreement with the experimental measurements.
\end{abstract}

\maketitle

Kerr optical frequency combs are obtained through pumping a high-$Q$ whispering-gallery mode (WGM) cavity with a resonant laser \cite{Delhaye_Nature_2007}. 
In the last decade, the experimental and theoretical study of these combs has permitted major advances in 
photonics (see review articles~\cite{Kipp_Science_2011, Chembo_Nanophot_2016, Lin_AOP_2017, Pasquazi_PR_2018}). 
From the applications standpoint, Kerr combs have been developed for time-frequency metrology, ultra-stable microwave generation, spectroscopy, and optical communications, just to name a few. From the fundamental perspective, Kerr combs have provided an ideal platform to investigate light-matter interactions in confined media.
It has been shown that a wide variety of dissipative structures could be excited in the WGM resonators, being either stationary (azimuthal roll patterns, cavity solitons, platicons) or non-stationary (breather solitons, spatiotemporal chaos, rogue waves). The primary bifurcations leading to these various patterns have also been the focus of a detailed analysis in the literature~\cite{Godey_PRA_2014,Parra,PPR_dark,Godey_EPJD_2017,Delcey_PTRSA_2018, Gelens, PPR_bright, PPR_patterns}. However, only a limited attention has been also devoted to secondary bifurcations, which lead to the destabilization of the stationary patterns \cite{ Perinet, PPR_patterns,Liu_secondary, Coulibaly}.

\begin{figure}[b]
\includegraphics[width=6cm]{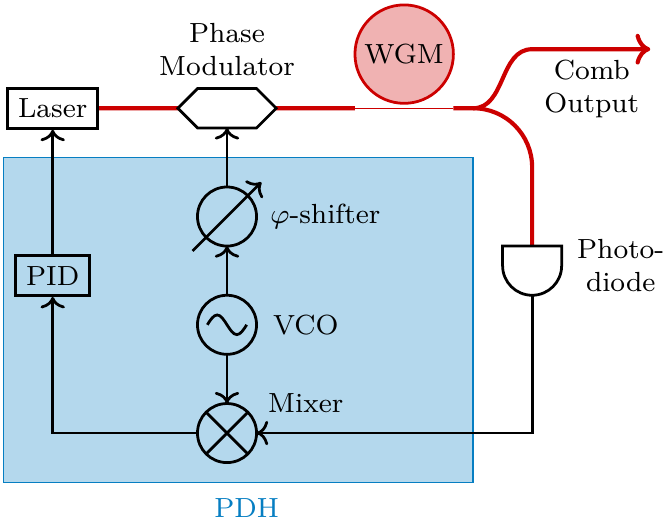}
\caption{(Color online) Experimental setup. 
PID: Proportional-integral-derivative controller;
VCO: Voltage-controlled oscillator;
PDH: Pound-Drever-Hall locking scheme;
WGM: Whispering-gallery mode resonator. 
}
\label{Setup}
\end{figure}

{ In this letter, we evidence experimentally one of these secondary bifurcations in 1D, namely the Eckhaus instability, which emerges when a roll (or stripes) pattern looses its stability against small-wavenumber perturbations.}
{ The Eckhaus instability has long been studied in fluid mechanics~\cite{Tuckerman,Ahlers}, liquid crystals~\cite{Lowe}, nonlinear optics~\cite{Louvergneaux01,Plumecoq,Li,Perinet,PPR_patterns}, or systems with delayed feedback~\cite{Wolfrum}. Experimental observations are however much more limited since large aspect-ratio patterns are required, while being difficult to attain in most systems. 
The Eckhaus instability can also be induced by spatial inhomogeneities \cite{Riecke,Plumecoq}, an effect that has been observed experimentally in a liquid crystal layer with optical feedback \cite{Louvergneaux01}.}
{ Other secondary instabilities and parametric perturbations may also hinder Eckhaus instabilities~\cite{Ahlers,Lowe}. Counterintuitively, despite their relatively small size, WGM resonators can output large aspect-ratio roll patterns with tens or even hundreds of peaks~\cite{Lin_OE_2015}, making the system more susceptible to develop small-wavenumber (or long-wavelength) instabilities}.

\begin{figure}[t]
\includegraphics[width=7cm]{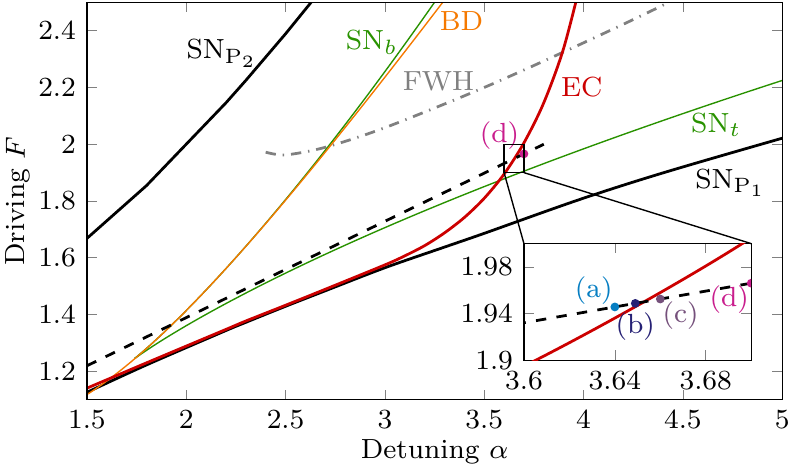}
\caption{(Color online) Bifurcation lines of the homogeneous solutions and the roll pattern with $L=55$ in the parameter space ($\alpha$, $F$). The HSS is stable below the MI line for $\alpha < 2$ and below the SN$_b$ line for $\alpha > 2$. The pattern is stable above the Eckhaus line (EC) and below the SN$_{P_2}$ or FWH line, whichever comes first. The dashed line shows the ramp of parameters applied to the pattern, starting from $(\alpha,  F)=(1 , 1.05)$ to $(3.8 , 2)$ beyond the Eckhaus instability.}
\label{phase_diagram}
\end{figure}

Our experimental system is displayed in Fig.~\ref{Setup}.
A MgF$_2$ WGM resonator with intrinsic quality factor $Q_\mathrm{in} = 1.8 \times 10^9$ is pumped by a resonant laser at 
$1552$~nm.   
The resonator has a diameter $d \simeq 11.8$~mm and group-velocity refraction index $n_\mathrm{g}=1.37$, yielding a free-spectral range $\mathrm{FSR}=c/n_\mathrm{g} \pi d \simeq 5.9$~GHz, where $c$ is the velocity of light in vacuum.
When the resonator is pumped above threshold, roll patterns emerging from  a Turing or modulational instability {(MI)} can be excited inside the cavity. They are characterized by an integer number of azimuthal rolls fitting the inner periphery of the disk. 
The integer number $L$ of rolls (or ``peaks'') in the azimuthal direction is the wavenumber of the pattern, and in the spectral domain, these roll patterns correspond to the so-called \textit{primary combs} where the teeth have a $L \times \mathrm{FSR}$ separation~\cite{Godey_PRA_2014,Parra,Sal16OE}.

\begin{figure}[t]
\includegraphics[width=7cm]{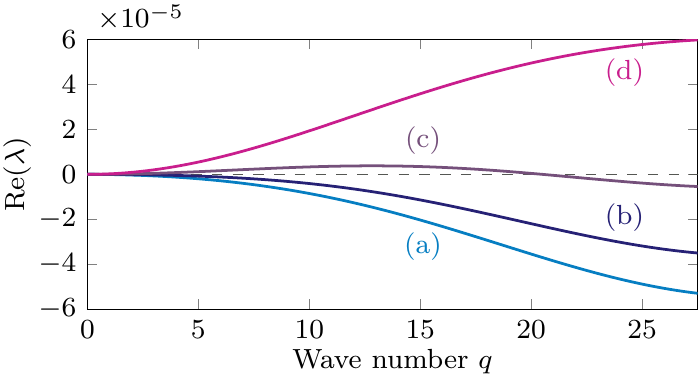}
\caption{(Color online) Real part of the eigenvalues of the pattern with $L=55$ for the  branch of soft modes obtained from Eq.~(\ref{eq_Upsilon}). Lines (a) to~(d) correspond to the parameter values indicated by blue dots in Fig.~\ref{phase_diagram}.
The curvature of the { branch} progressively changes from negative to positive { signaling} the Eckhaus instability.}
\label{lambdaq}
\end{figure}

The theoretical analysis of the Eckhaus instability starts with the Lugiato-Lefever equation~\cite{Lugiato}, 
which is an accurate model to analyze the laser field dynamics in Kerr-nonlinear WGM 
resonators~\cite{Matsko_OL_2011, Yanne_PRA_2013, Coen_OL_2013}.
The slowly varying complex amplitude of the normalized intracavity field $\psi(\theta,\tau)$ obeys the equation  
\begin{eqnarray}
\label{LLE}
\frac{\partial \psi}{\partial \tau} = -(1 + i \alpha) \psi - i \frac{\beta}{2} \frac{\partial^2 \psi}{\partial \theta^2} 
                                       + i |\psi|^2 \psi + F \, ,
\end{eqnarray}
where $\theta \in [-\pi,\pi]$ is the azimuthal coordinate along the ring of the
resonator, and { $\tau= t/2\tau_{\rm ph}$ is the time scaled to the photon lifetime}. The normalized parameters of this equation are 
the continuous-wave pump field $F$,
the frequency detuning between laser and pumped { resonance} frequencies $\alpha$,
and the group-velocity dispersion $\beta$~\cite{Yanne_PRA_2013}.

Equation~(\ref{LLE}) has homogeneous steady states $\psi_{\rm s}$ implicitly given by
$\rho_{\rm s}[1+(\rho_{\rm s} - \alpha)^2]=F^2$ with $\rho_{\rm s}=|\psi_{\rm s}|^2$. 
{ The solution is trivalued for $\alpha > \sqrt{3}$. The line SN$_{b}$ (resp. SN$_{t}$) in Fig.~\ref{phase_diagram} corresponds to the saddle-node bifurcation where lower (resp. upper) and middle branches meet, so that SN$_{b}$ and SN$_{t}$ unfold from cusp at $\alpha = \sqrt{3}$~\cite{Godey_PRA_2014,Parra}. For $\alpha < \sqrt{3}$ the solution is monovalued. In what follows we refer the lower homogeneous steady state as HSS.}

In the anomalous regime ($\beta<0$) { and for $\alpha < 2$,  $\rho_{\rm s}=1$ is the MI threshold  above which the HSS is unstable to perturbations with wavenumber $L$ in the neighborhood of $L_\mathrm{u}=\sqrt{(2/\beta)(\alpha -2\rho_{\rm s})}$ (see Fig.~\ref{phase_diagram}). Roll patterns with different wavenumbers can emerge although typically the one with wavenumber $L_\mathrm{u}$ dominates since it has the largest growth ratio.} This pattern is supercritical for $\alpha<41/30$, and sub-critical for $\alpha > 41/30$. { Regarding the other possible roll patterns, it turns out that only those with wavenumber close to $L_\mathrm{u}$ are}
stable, forming what is known as a \textit{Busse balloon}~\cite{Walgraef,Cross} while the others are unstable.  Moreover, in the subcritical regime, cavity solitons or localized states (LSs) coexist with the periodic patterns and the HSS. For $\alpha >2$ the critical wavenumber is zero and the { threshold} $\rho_{\rm s}=1$ { is} a Belyakov-Devaney (BD) transition of the HSS~\cite{Parra,PPR_patterns} { (see Fig.~\ref{phase_diagram})}.

\begin{figure}
\includegraphics[width=8cm]{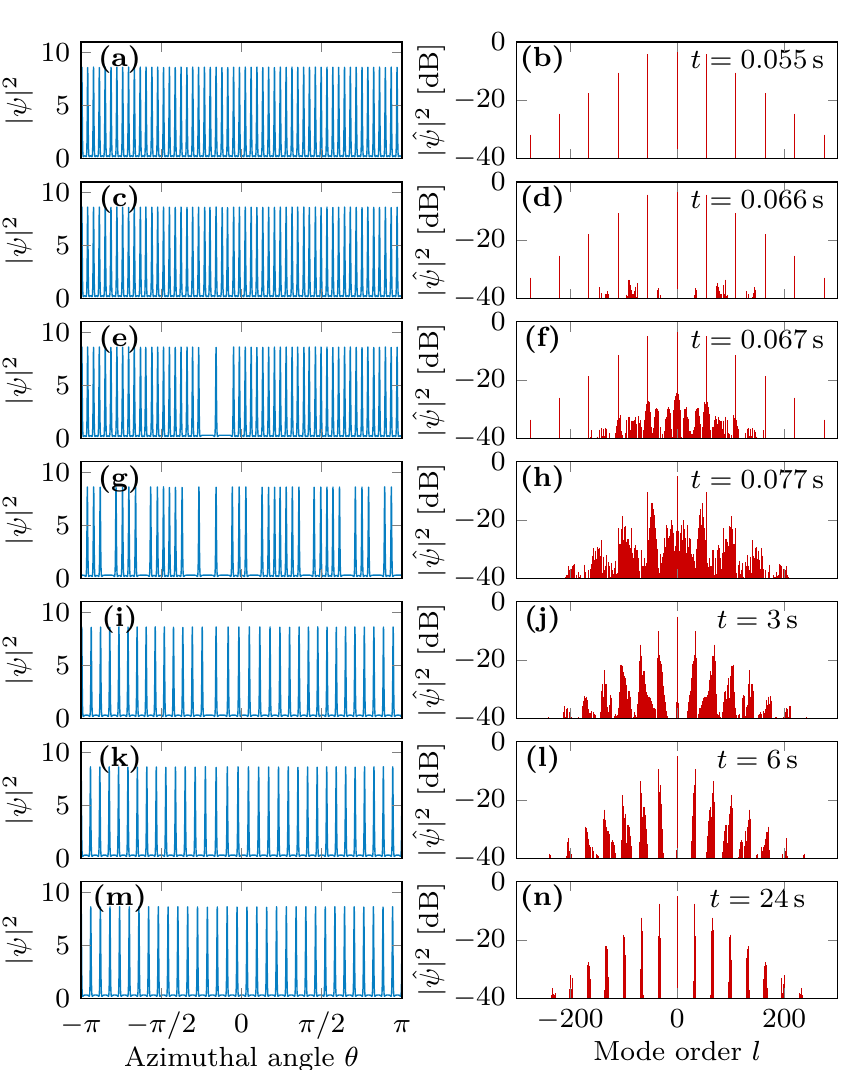}
\caption{(Color online) Numerical evidence of Eckhaus instability. Nonlinear evolution of the $L=55$ pattern after crossing the EC line as described in the main text. Left column shows the spatial profile $|\psi(\theta)|^2$ of the pattern at different times, while right column shows the corresponding power spectra  $|\hat{\psi}(l)|^2$. Time stamps given in real time $t= 2 \tau_\mathrm{ph} \times \tau = 10^{-6} \tau$. These numerical simulations show that after the Eckhaus bifurcation, the convergence towards the new pattern $L=33$ is a very slow process that takes place in the timescale of a minute. The time needed to simulate this $24$~s-long transient dynamics was about one month, using a pseudospectral algorithm where the linear terms in Fourier space are integrated exactly while the nonlinear ones are integrated using a second-order approximation in time \cite{footnote2}.}
\label{fig_Eckhaus}
\end{figure}

To study the secondary bifurcations that destabilize a roll pattern of wavenumber $L$ we perform a linear stability analysis.
The stationary but $\theta$-dependent pattern can be { expanded in Fourier series}
\begin{equation}\label{sta_ansatz2}
\psi_{_\mathrm{P}}(\theta) = \displaystyle\sum_{n=-N}^{N{-1}} \psi_n e^{i n L \theta},
\end{equation}
with $L$ being the integer wavenumber (or order) of the pattern { and} $\psi_n$ the complex amplitudes of the Fourier modes. We take $N=32$ and the amplitudes can be calculated numerically by solving the stationary problem using a Newton-Raphson algorithm. 
Linearizing Eq.~(\ref{LLE}) about the stationary pattern $\psi_{_\mathrm{P}}(\theta)$ yields the perturbation equation 
\begin{eqnarray}\label{linear_complex}
\partial_\tau \delta \psi &=& -(1+i\alpha) \delta \psi-i(\beta/2)\partial_\theta^2 \delta \psi  \nonumber \\
                        && +2i|\psi_{_\mathrm{P}}|^2\delta \psi+i\psi_{_\mathrm{P}}^2\delta \psi^* \, .
\end{eqnarray}
Due to the periodicity of the { system}, the solution of { Eq.~(\ref{linear_complex}}) can be written as the superposition of Bloch waves 
\begin{equation}\label{eq_def_del_psi}
\delta \psi(\theta, \tau)=e^{iq\theta} \delta a(\theta,\tau,q)+e^{-iq\theta}\delta a(\theta,\tau,-q) \, ,
\end{equation}
where $\delta a$ has the same periodicity of the pattern $\psi_{_\mathrm{P}}(\theta)$, and can be written as
\begin{equation}\label{sta_ansatz1}
 \delta a(\theta,\tau,q)=\displaystyle\sum_{n=-N}^{ N{-1}}  \delta a_n (\tau,q)e^{inL \theta} \, 
\end{equation}
with $q$ being an integer.
Using Eq.~(\ref{linear_complex}), a set of linear equations for the Fourier modes $\delta a_n(\theta,q)$ can be 
derived~\cite{Gomila07}, and in compact form they read as 
\begin{equation}
 \partial_\tau \Upsilon(\tau,q)=\mathbf{\mathsf{M}}(\{\psi_n\},q) \Upsilon(\tau,q) \, ,
\label{eq_Upsilon} 
\end{equation}
where $\Upsilon(\tau,q) \equiv [\delta a_{-N} (\tau,q),\cdots,\delta a_{N{-1}}(\tau,q),\delta a^*_{-N}(\tau,-q),\\\cdots,\delta a^*_{N{-1}}(\tau,-q)]$.
The stability analysis of $\psi_{\mathrm{P}}(\theta)$ reduces to find the $2N$ eigenvalues $\{\lambda_n(q)\}$ of the 
matrix $\mathbf{\mathsf{M}}(\{\psi_n\},q)$, and its corresponding eigenvectors, for each value of $q$. 
The eigenvalues for a given integer $q$ determine the stability of the pattern against perturbations containing  any set of wavenumbers $nL \pm q$. For this analysis it is sufficient to consider only the $q$ values inside the 
first Brillouin zone $[0,L/2]$.
{ We recall that $q=0$ corresponds to the Goldstone mode associated to the translational invariance, and modes with $q \gtrsim 0$  form the branch of soft modes}. 

\begin{figure}[t]
\includegraphics[width=6cm]{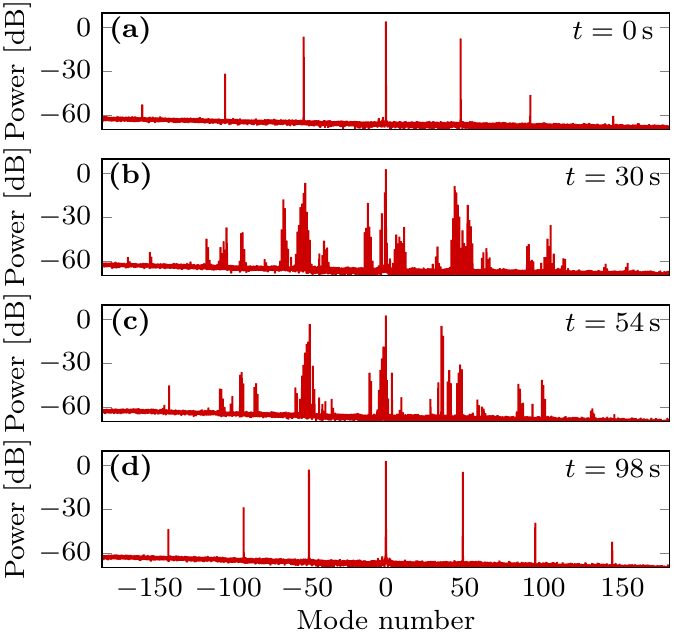}
\caption{(Color online) Experimental evidence of ultra-slow Eckhaus instability around a roll patterns of wavenumber $L=50$.
The pattern becomes Eckhaus-unstable with a temporal evolution characterized by minute timescale transients, before converging towards another pattern of lower order $L=47$.  }
\label{experimental_combs}
\end{figure}

{ Fig.~\ref{phase_diagram} shows the bifurcation lines of the roll pattern created spontaneously with the most unstable wavenumber $L_\mathrm{u}$ for $\alpha=1$ and $F=1.05$ ($\rho_{\rm s}=1.095$). For the value $\beta=-8 \times 10^{-4}$ considered here, we have $L_\mathrm{u}=55$. As the detuning is increased, the pattern becomes subcritical for $\alpha \simeq 41/30$, and above this value it exists between the saddle-node lines SN$_{\rm{P}_1}$ and SN$_{{\rm P}_2}$, { although unstable below the Eckhaus line (EC). 
Above a certain value of the detuning and the pump, we observe a finite-wavelength Hopf (FWH) instability (dot-dashed) leading to oscillatory patterns~\cite{Gomila07,PPR_patterns}.}
We will not consider this regime { here since we} focus on the Eckhaus instability.}
{ Note that pattern and HSS are stable and coexist in the parameter region limited by MI, SN$_b$, FWH and EC lines.}

Figure~\ref{lambdaq} shows the real part of the eigenvalues of the pattern as a function of the wavenumber $q$ { for the  branch of soft modes \cite{Gomila07}.}
The parameters correspond to those of the blue dots in Fig.~\ref{phase_diagram}, while crossing the Eckhaus instability.
The change of convexity of the branch { at} $q=0$ is what precisely signals the Eckhaus instability. { After the instability}, the pattern becomes unstable to small-wavenumber perturbations. Well { beyond} the instability, the { mode with maximum growth rate has a wavenumber close to} the edge of the Brillouin zone $q=L/2$.

After encountering an Eckhaus instability a pattern with a wavenumber which is too large to be stable looses cells in such a way that the new wavenumber lies in the stability balloon~\cite{Perinet}. For supercritical patterns this happens at a relatively fast time scale.
For subcritical patterns, the HSS is stable and coexists with the pattern allowing the formation of LS. When a cell is lost the space is occupied by the HSS leading to a transient state formed by groups of LSs separated by the HSS, known as cracking pattern~\cite{Harkness}.
If LSs have oscillatory tails they may lock at specific distances given by multiples of the oscillatory tail wavelength, thus the cracking pattern is stationary. On the contrary, if LS tails are monotonous, LS repel each other and the cracking pattern evolves 
towards a periodic solution with equally spaced peaks and a stable wavenumber. In practice, a similar behavior is observed if tails are oscillatory with a wavelength much larger than the typical separation between peaks.
This transient behavior can be extremely slow as the interaction decays exponentially with the distance between peaks allowing for long-lived cracking patterns likely to be observed at second- and even minute-timescale in experiments. 

In our numerical simulations, the Eckhaus instability is triggered by slowly ramping up the detuning { and the pump} parameter.
This procedure is consistent with the experimental system where the detuning is thermally driven across the 
resonance~\cite{Dia15OL}. 
The dotted line in Fig.~\ref{phase_diagram} shows the ramp of parameter values used in the simulation shown in 
Fig.~\ref{fig_Eckhaus}. The simulation starts at $t = 0$ with $\alpha=1$ and $F=1.05$,   just above the MI, and a stable pattern with $L=55$ emerges, corresponding to the wavenumber with maximum growth rate $L_\mathrm{u}$. The parameters are ramped until $t=0.05$ s with $\alpha=3.8$ and $F=2$, above the Eckhaus instability. The simulation then continues up to $t=24$~s (which corresponds to {24}~million photon lifetimes in our resonator), with clamped values for $\theta$ and $F$. The original pattern, whose spatial profile and power spectrum is shown Figs.~\ref{fig_Eckhaus}(a) and~(b), remains stable through the ramp until it crosses the EC line. At this point  the pattern becomes unstable and soft mode perturbations start to grow. As a consequence some pattern cells disappear as shown in 
Figs.~\ref{fig_Eckhaus}(c) and~(d). Further
development of the instability leads to a cracked pattern { as shown in 
Figs.~\ref{fig_Eckhaus}(e) and~(f) for time $t=0.077$~s. For the parameters considered, LSs have oscillatory tails although the wavelength of the tail oscillations is much larger than the separation between consecutive peaks \cite{footnote}.
As a consequence, LSs do not get pinned, but they repel each other instead}. This ultra-slow dynamics can take more than a minute to converge asymptotically to another pattern, as observed experimentally in Fig.~\ref{experimental_combs}.
In contrast for all nonlinear effects reported so far using the LLE, the transient dynamics usually last only few $\tau_\mathrm{ph}$ (i.~e., few~$\upmu$s in our case). 
Note that if the ramp is increased to much larger values of the detuning, one reaches the single-soliton regime described in \cite{Herr} and eventually only a single peak survives.

Figure~\ref{experimental_combs} shows an experimental example { of} primary comb of order corresponding to a high-wavenumber roll patterns { with} $L=50$. When the laser frequency is thermally driven across from the resonance, we observe the emergence of spurious peaks around the main primary comb, and the comb dynamics is characterized by a very slow timescale, that can be larger than a  minute.
This timescale appears \textit{a priori} as inconsistent with the intrinsic Kerr comb dynamics, where the slowest timescale is generally the photon lifetime $\tau_\mathrm{ph}=Q/\omega_0 \sim 1$~$\upmu$s, with $Q \sim 10^9$ being the loaded quality factor of our resonator and $\omega_0$ is the angular frequency of the pumped mode~\cite{Che10OL,Coi13JOVE,Hen15OL}. 
However, as demonstrated earlier, a detailed analysis unveils that the mechanism { behind} this ultra-slow timescale dynamics is {an} Eckhaus instability leading to very-long-lasting transient cracking patterns, and later on to a lower order stable roll pattern.

In conclusion, we have experimentally evidenced Eckhaus instability in a whispering-gallery mode resonator. 
The emerging timescale of the instability, dominated by the interaction between LSs, was shown to be six to eight orders of magnitude larger than the intracavity photon lifetime, which is the natural timescale for Kerr comb dynamics. These results permit to achieve a deeper understanding of secondary bifurcations in dissipative optical systems, and future work will investigate in detail the wide variety of spatiotemporal patterns that can be excited via these bifurcations, including when co-induced by other bulk nonlinearities~\cite{Lin_2016_Raman, Lin_2016_Univ, Lin_AOP_2017}. \\

D.~G. and Y.~K.~C would like to acknowledge support from the project ND-PHOT jointly funded by CSIC and CNRS. 
D.~G. and P.C. acknowledge  financial  support  from  Agencia  Estatal  de  Investigaci\'on
(AEI, Spain) and Fondo Europeo de Desarrollo Regional under Project SuMaEco, grant number: RTI2018-095441-B-C22 (AEI/FEDER,UE) and Agencia Estatal de Investigaci\'on through Mar\'ia de Maeztu Program for Units of Excellence in R\&D (MDM-2017-0711). Y.~K.~C. also acknowledges funding from the European Research Council through the projects NextPhase \& Versyt, from the \textit{Centre National d'Etudes Spatiales} (CNES) through the project SHYRO, and from the University of Maryland.


\begin{thebibliography}{99}
\bibitem{footnote2} The asymptotic numerical error is a numerical noise floor that is independent of the length of the simulation, since we are simulating the convergence of a system towards a stable fixed point (attractor) in the state space.
\bibitem{footnote} The wavelength of the spatial oscillations in the tails of the LS calculated from the imaginary part of the spatial eigenvalues of the HSS \cite{Parra17} for the parameters used in the simulations is $\pi/6.59$, larger than the typical separation between peaks.
\end{thebibliography}
\end{document}


\title{SUPPLEMENTAL MATERIAL \\
       Observation of the Eckhaus Instability in Whispering-Gallery Mode Resonators}

\author{Dami\`a Gomila}
\author{Pedro Parra-Rivas} 
\author{Pere Colet} 
\author{Aur\'elien Coillet} 
\author{Guoping Lin} 
\author{Thomas Daugey} 
\author{Souleymane Diallo} 
\author{Jean-Marc Merolla} 
\author{Yanne K. Chembo}

\noaffiliation

\date{\today}

\begin{abstract}
This supplemental material provides additional information in relation to the whispering-gallery mode resonator (properties, polishing process, cavity-ring-down characterization of the quality factor), to the Pound-Drever-Hall (PDH) stabilization scheme, and to the numerical simulations of the ultra-slow transients subsequent to the Eckhaus instability.   
\end{abstract}

\maketitle

\section{The whispering-gallery mode resonator: properties, polishing process, and characterization}

Magnesium fluoride ($\rm MgF_2$) is a tetragonal birefringent crystal with a transparency window ranging from UV to the mid-infrared window ($0.1$--$8$~$\upmu$m). It is used as material for the fabrication of lenses and polarizers. It has also been widely used for Kerr comb experiments at telecom wavelength. Its transparency at $1550$~nm makes it suitable for the fabrication of millimetric ultra-high-$Q$ whispering-gallery mode resonators at the mentioned wavelength. Its refractive index is $1.3717$ at $1550$~nm, and also makes it suitable for taper coupling. From a commercially available crystalline $\rm MgF_2$ WGM disk, consecutive steps of grinding and polishing allows us to achieve a quality factor of $\sim 10^9$, using an air-bearing spindle motor to spin the disk. During the grinding step, we used decreasing abrasive-coated size support to shape the resonator rim into a sharp ``V" where the whispering-gallery mode will propagate. The second step consists in fine-grain polishing of the resonator via decreasing size abrasive particles, down to $100$~nm. Once the final polishing step is done (surface roughness below $5$~nm), we proceed to the characterization of the resonator in order to obtain its intrinsic, extrinsic and loaded quality factor.

\begin{figure}[b]
	\centering
	\includegraphics[width=6cm]{Suppl_MgF2_CRD}
	\caption{Ring down transmission spectrum. A theoritical fit gives an intrinsic quality factor of $ \rm 1.82 \times 10^9$, an extrinsic quality factor of $ \rm 4.33 \times 10^9$ and a loaded quality factor of $\rm 1.28\times 10^9$}
	\label{fig:MgF2_CRD}
\end{figure}

We have used the cavity-ring-down measurement technique to characterize the resonator quality factor. It is an efficient technique that allows to avoid thermal effects but permits nevertheless to obtain intrinsic, extrinsic and loaded quality factors. A continuous-wave laser with sub-kHz linewidth swept at a scanning speed of $1.2$~GHz/ms is used to couple light into the resonator through the evanescent field of a tapered silica fiber. The recorded transmission is the temporal interference pattern between the laser input and the decaying resonance light. As shown in Fig.~\ref{fig:MgF2_CRD}, a fitting of the experimental data gives an intrinsic quality factor of $ \rm 1.82 \times 10^9$, an extrinsic quality factor of $ \rm 4.33 \times 10^9$ and a loaded quality factor of $\rm 1.28\times 10^9$.

\begin{figure*}[t]
\begin{center}
\includegraphics[width=14cm]{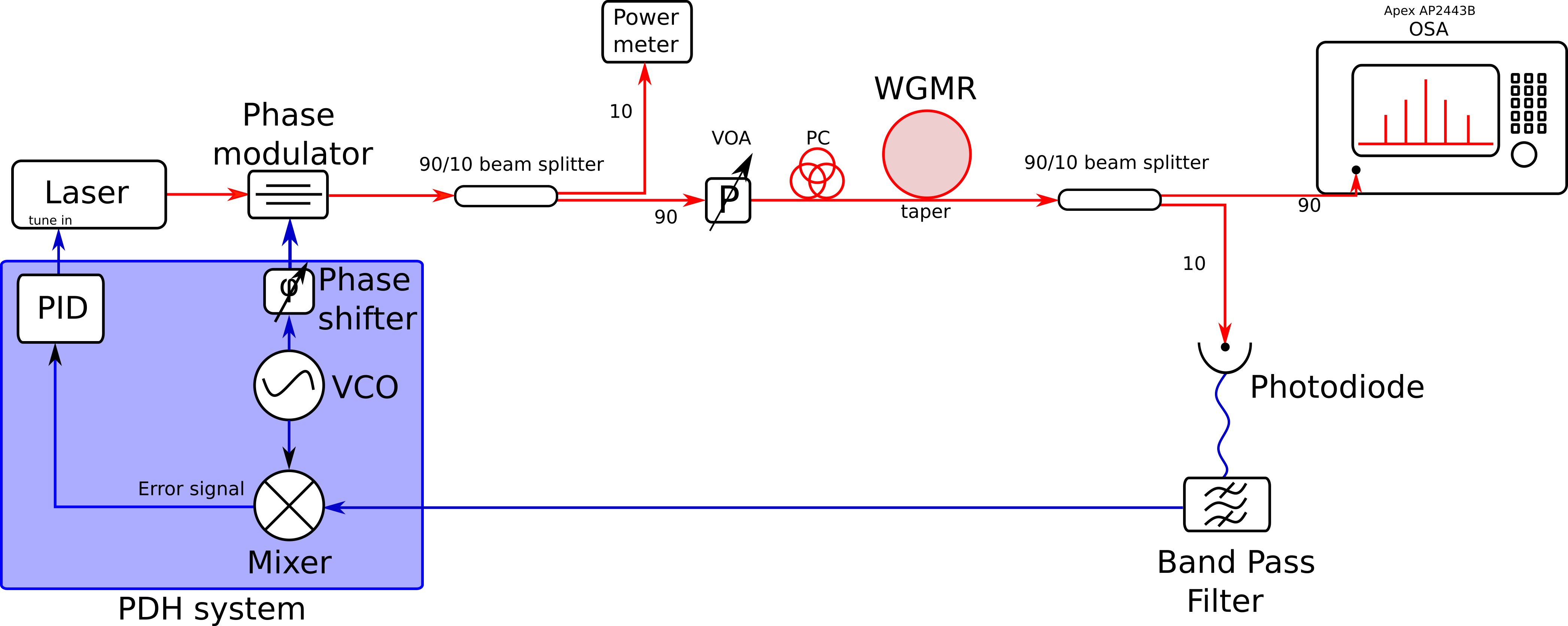}
\end{center}
\caption{Experimental setup used to generate frequency combs with a whispering-gallery mode resonator. Blue lines represent electrical paths whereas red lines represent optical paths.
\label{fig:experimental_setup}}
\end{figure*}

\section{Pound-Drever-Hall stabilization system}

The PDH technique is designed to stabilize the pump laser wavelength to a particular resonance.
It was initially used to improve laser frequency stability by locking it on the
mode of a cavity (more stable that the laser itself), thus transferring the stability of the cavity mode frequency to the laser. It also has the effect of improving the linewidth of the locked laser if the mode linewidth is narrower than that of the laser.
It has found many applications in very different fields, from the detection of gravitational
waves  to high-resolution spectroscopy.
In our case, the PDH locking technique is not used to improve the stability of the laser but rather to actively lock the laser onto an thermally unstable resonance of the whispering-gallery mode resonator. 
 	
WGM resonators are extremely sensitive to temperature and when light is injected inside a resonance, a part of the optical power is absorbed, causing the resonator to heat up. This effect both changes the refractive index and increases the diameter of the resonator via thermal dilatation, causing the resonances to shift. In order to control these effects, we actively lock the laser onto the resonance so that the laser follows the resonance as it shifts.
We use a Toptica PDH module which generates the error signal, handles all communications with a computer, and drives the high-voltage source ($0$--$150$~V) needed to control the piezoelectric element in the external cavity of our laser (NKT adjustik).

The experimental setup is represented on Fig.~\ref{fig:experimental_setup}.
It features a narrow-linewidth  ($<1$~kHz) continuous-wave laser operating around $1550$~nm, which can be swept over $15$~pm by applying a voltage on the piezoelectric element of the external cavity of the laser. A phase modulator is used in conjunction with the Toptica PDH module to phase-modulate the laser and obtain an error signal. The laser then goes through a $90/10$ optical coupler, with $10\%$ of the power being sent to a powermeter to have an image of the input power in the resonator, while the remaining $90\%$ is sent in a polarization controller (PC) and then through a tapered fiber.   

A WGM is coupled to the fiber, by sweeping the laser. We can observe the different coupled modes inside the resonator and the polarization controller allows us to optimize the coupling of those modes. The WGM optical resonator that was used for this experiment is a $12$~mm diameter MgF$_2$ disk, which was mechanically shaped and polished from a commercially available preform as explained in the preceding section. The output of the resonator is then split into two path, one leading the photodiode and the PDH module and the other path leading to a high resolution optical spectrum analyzer (Apex AP2443B).

The experimental procedure is the following. The laser is scanned on the largest span possible. Depending on the coupling we will observe a certain number of modes. Among those modes, we select the ones with a high coupling and high quality factor. The span is then reduced in order to sweep over only one mode. When Kerr comb generation is confirmed by the OSA, the laser is then locked to that mode with the PDH module.

\section{Numerical simulations and movie of the Eckhaus instability}

The numerical simulation of the Lugiato-Lefever Equation (LLE) is done using a pseudospectral method where the linear terms in Fourier space are integrated exactly while the nonlinear ones are integrated using a second-order approximation in time. A 1D lattice of $4096$ points was used. The time $t$ in seconds is expressed as $t= 10^{-6} \tau$, where $\tau$ is the dimensionless time of the LLE. The time-step for numerical integration was $\Delta \tau=0.001$. 

The movie added to this Supplementary Material shows the time evolution of the cracking patterns, corresponding to Fig.~4 of the main document. The upper panel shows the intracavity field in the rotating frame while the bottom panel shows its corresponding frequency comb (Fourier transform). The time and the current value of the detuning are displayed on the top right corner of the bottom panel. The simulation starts a $\tau=0$ with an homogeneous initial condition with small random noise, the pump $F=1.05$ and detuning $\alpha=1$ (just above threshold). A stable frequency comb is rapidly formed. Meanwhile the pump and detuning are progressively increased following a ramp (see text) until $\tau=50000$ with $F=2$ and $\alpha=3.8$. For this values of the parameters the initial comb is Eckhaus unstable and side band modes start to growth. The nonlinear evolution of the EI switches off several peaks leading to a cracking pattern around $\tau=77000$. This pattern evolves at a much slower time scales as the localized structures interact and move apart giving rise to a new frequency comb with a stable wavenumber in the very long run.